\documentclass[aps, twocolumn]{revtex4}

\usepackage{float}
\usepackage{dcolumn}
\usepackage{amsmath}
\input{epsf}
\input{epsfx}

\ifx\pdftexversion\undefined
  \usepackage{graphicx}
\else
  \usepackage[pdftex]{graphicx}
\fi
\usepackage{epsfig}
\usepackage{ulem}

\begin{document}
\title{Local Quantum Dot Tuning on Photonic Crystal Chips}

\author{Andrei Faraon \footnote[1] {Electronic address: faraon@stanford.edu }, Dirk Englund, Ilya Fushman, Nick Stoltz \footnote[2] {Also at Department of Electrical and Computer Engineering, University of California Santa Barbara, CA 93106.}, Pierre Petroff \footnotemark[2], Jelena Vu\v{c}kovi\'{c}}
\affiliation{E. L. Ginzton Laboratory, Stanford University, Stanford, CA, 94305}

\date{March 26, 2007}

\begin{abstract}
 Quantum networks based on InGaAs quantum dots embedded in photonic crystal devices rely on QDs being in resonance with each other and with the cavities they are embedded in. We developed a new technique based on temperature tuning to spectrally align different quantum dots located on the same chip. The technique allows for up to $1.8 nm$ reversible on-chip quantum dot tuning.

\end{abstract}

\maketitle

	Solid-state approaches to quantum information have generated tremendous interest over the past decade.  Recent schemes often employ cavity quantum electrodynamics (CQED) to manipulate qubits, with many using quantum dots (QDs) coupled to optical cavities.  Cavities in photonic crystals (PCs) are particularly attractive due to their small mode volume and high quality factor \cite{05PRLEnglund, Yoshie04}.  In addition, photonic crystals are ideal for integrating devices into an on-chip network for information processing \cite{CZKM1997PRL, DirkAndreiSPhtransfer}.  A major problem facing such proposals, however, is spatial and spectral matching of distinct inhomogeneously broadened quantum dots.  Spatial alignment can be achieved either by positioning the PC cavity on an already identified QD \cite{HennessyDeterministicCoupling, SCImamogluNature, PhotolithQDRegistration}, or by relying on chance. 	For spectral alignment, there are a few techniques that can be used to modify the emission wavelength of InGaAs quantum dots: Stark shift \cite{StarkKarraiPRL}, Zeeman shift \cite{ZeemanKarraiPRL}, temperature tuning \cite{TTunePetroffAPL} and strain tuning \cite{InGaAsStrainTune}. In this paper, we demonstrate a technique for independent control of QDs, employing structures with high-$Q$ cavities whose temperature is controlled by laser beams.  We discuss the thermal and optical design, device fabrication, and testing.  Our in-situ technique allows extremely precise spectral tuning of InGaAs quantum dots by up to 1.8 nm and of cavities of up to 0.4 nm (4 cavity linewidths).  The technique is crucial for spectrally aligning distinct quantum dots on a photonic crystal chip and forms an essential step toward creating on-chip quantum information processing devices.

		To achieve independent on-chip tuning, distinct regions containing the quantum dots of interest must be kept at different temperatures. Since GaAs is a good thermal conductor, on-chip local thermal insulation must be provided to achieve significant local heating. For this reason, we fabricated suspended PC structures with minimal thermal contact to the rest of the chip (Fig.1(a)). The suspended PC structures were fabricated on a quantum dot wafer grown by molecular beam epitaxy on a Si n-doped GaAs (100) substrate with a $0.1 \mu m$ buffer layer. To increase quantum dot collection efficiency, a 10 period distributed Bragg reflector (DBR) mirror underneath the QDs is included, consisting of alternating layers of AlAs/GaAs with thicknesses of 80.2/67.6 nm respectively . A 918 nm sacrificial layer of Al$_{0.8}$Ga$_{0.2}$As is located above the DBR mirror. The active region consists of a 150 nm thick GaAs region with a centered InGaAs/GaAs QD layer. QDs self-assemble during epitaxy operating in the Stranski-Krastanov growth mode. InGaAs islands are partially covered with GaAs and annealed before completely capping with GaAs. This procedure blue shifts the QDs emission wavelengths towards the spectral region where Si-based detectors are more efficient.

\begin{figure}[htbp]
    \includegraphics[width=3.375in]{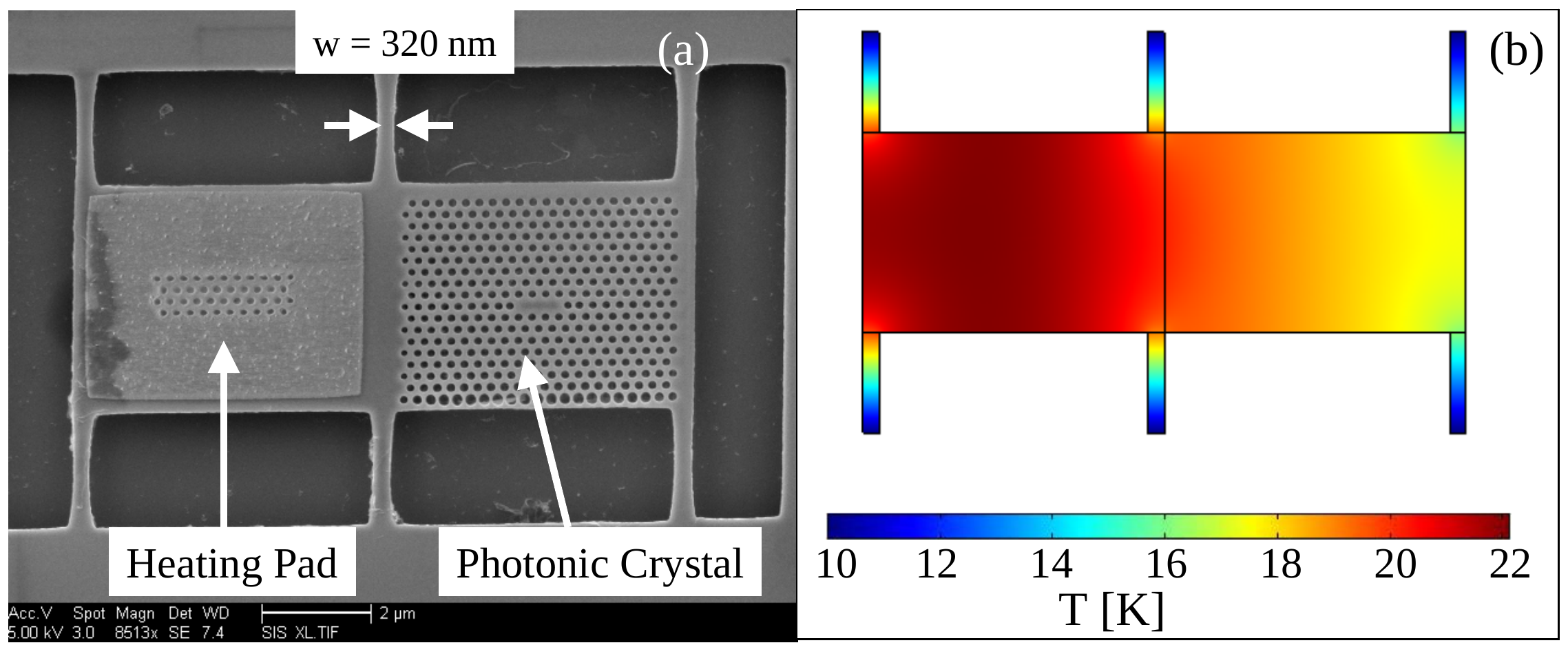}
    \caption{(a) Scanning electron microscope image of the fabricated structure showing the PC cavity, the heating pad and the connection bridges. The temperature of the structure was controlled with a laser beam (960 nm) focused on the heating pad. (b) Simulation of the temperature distribution in the device for $10^{-2} mW$ absorbed in the heating pad. This is only an order of magnitude estimation of the temperature distribution, because the thermal conductivity changes significantly as a function of temperature and local geometry.}
    \label{fig1}
\end{figure}

	The fabricated structures (12 $\mu m$ long, 4 $\mu m$ wide, 150 $ nm$ thick) consist of a PC cavity and a heating pad (Fig.1 (a)). To provide the thermal insulation needed for efficient device heating, the structure was connected to the rest of the chip by only six narrow bridges. The thermal conductivity of narrow ($ \approx 100nm$), cold (4K - 10K) GaAs bridges is reduced by up to four orders of magnitude with respect to the bulk GaAs \cite{FonGaAsThermalCond}, thus improving the thermal insulation. We tested two devices with connection bridges of the same length (2 $ \mu m $) but different widths: $w = 320 nm$ and $w = 800 nm$. The temperature of the device was controlled by using a focused laser beam to heat up the pad next to the photonic crystal cavity. To minimize background photoluminescence in single quantum dot measurements, the heating laser is tuned below the QD absorption frequency. A metal layer (20 nm Cr/15 nm Au) was deposited on the heating pad to increase heat absorption. The thermal conductivity of GaAs beams with cross sections on the order of $100 nm$/$100 nm$, and the absorption coefficient of the metal layer are not well known. As measured by Fon et. al \cite{FonGaAsThermalCond}, the thermal conductivity of GaAs beams with dimensions $100nm$/$200nm$/$6 \mu m$ is about $3 \times 10^{-2} W K^{-1} cm^{-1}$ at $10 K$, three orders of magnitude lower than the bulk value. Because of the size similarity, we assume that the connection bridges from our device have a similar thermal conductivity. Assuming that $10 ^{-2} mW$ of heat is absorbed in the heating pad and considering that the device has the same thermal conductivity as the bridges, we simulated the temperature distribution inside the cantilever. The simulation indicates that the photonic crystal region will be heated up to $\approx 20 K$ (Fig.1 (b)). However, this simulation gives only an order of magnitude estimation of the temperature distribution, because the thermal conductivity changes significantly as a function of temperature and local geometry.

	The local quantum dot tuning measurements were performed in a continuous flow liquid helium cryostat maintained at 10K. A Ti:Saph laser tuned at 855 nm was used to excite the quantum dots while a 960 nm laser diode acted as the heating laser. Using a pinhole we collected photoluminescence from a quantum dot located inside the photonic crystal slab. By increasing the power of the heating laser, QD emission was observed to redshift (Fig.2 (a)). The QD line-width broadens with increasing heating pump power, as expected from experiments where the full sample is heated. We were able to tune the quantum dot by 1.4 nm while the linewidth broadened from $0.04 nm$ to $0.08 nm$. The quantum dot could be further shifted by 1.8 nm but the PL intensity dropped rapidly. To show the compatibility of this local tuning technique with single photon measurements and quantum information processing, we proved anti-bunched single exciton emission using a Handbury-Brown-Twiss interferometer while the emission line was shifted by $ 0.8 nm$ (Fig.2 (b)).

\begin{figure}[htbp]
    \includegraphics[width=3.375in]{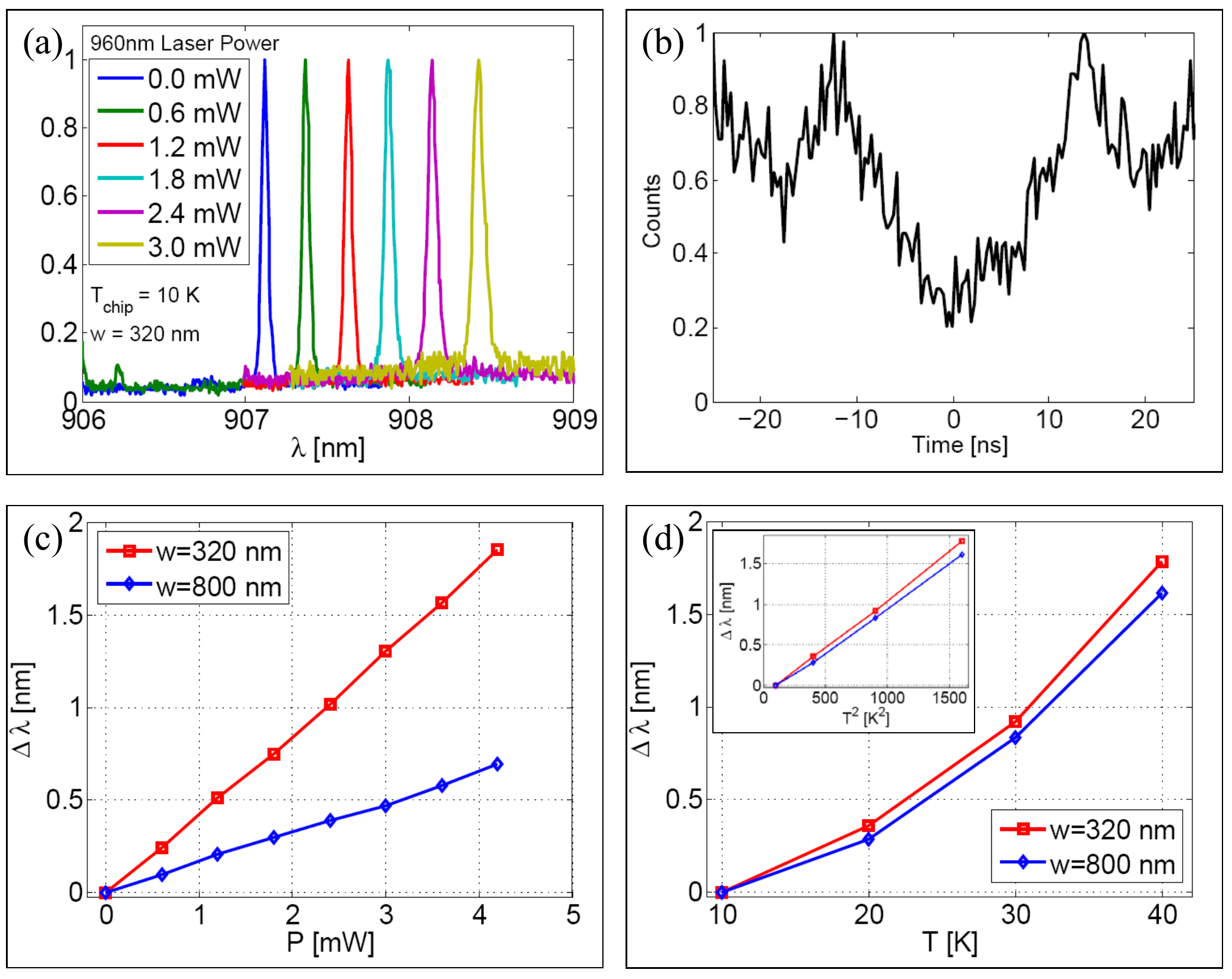}
    \caption{(a) Quantum dot tuning vs. heating pump power. The structure is connected to the substrate by bridges measuring $320 nm$ in width. The quantum dot emission shifts by 1.4 nm while increasing the heating laser power to 3 mW. Only a small fraction of the heating laser power is absorbed in the metal pad. (b) Autocorrelation measurement showing single photon anti bunching while the QD was detuned by 0.8 nm using the local tuning technique. (c) Dependence of the QD detuning on the heating laser power. The two data sets correspond to structures with different thermal contact to the substrate ($320nm$ and $800 nm$ bridges). (d) QD temperature tuning by changing the temperature of the entire chip by heating the cryostat. The inset shows that the detuning is linear in $ T^2 $}
    \label{fig2}
\end{figure}

	To investigate the thermal properties of the fabricated devices, we compared the shift of quantum dots located on structures with different bridge widths, $w = 320 nm$ and $w = 800 nm$. The thermal conductance of the bridges is proportional to their width $ w $. Under the same pump conditions the temperature of the structure is inversely proportional to $w$ so we would expect the QD to shift  2.5 ( i.e. 800/320 ) times further for the structure with thinner bridges. The QD shift observed on the two structures is plotted in Fig.2 (c). For the same pump power, the QD shifts 2.65 times further for $w = 320nm$ than for $w = 800 nm$, in good agreement with the expected result.

	The temperature dependence of the QD shift was determined by changing the temperature of the entire chip by heating the cryostat. The results are plotted in a Fig.2 (d) and indicate that a shift of 1.8 nm corresponds to a temperature of 40 K. This implies that during the local temperature tuning experiment the structure was also heated up to 40 K. The QD shift shows a quadratic dependence with temperature (Fig.2 (d) inset), which is expected since the band gaps of GaAs/InGaAs have a quadratic temperature dependence in this temperature interval \cite{GaAsBandGapThurmond}. Our experimental data shows a linear dependence of the QD shift with the heating laser power (Fig.2(c)), which implies a linear relation between the power of the heating laser and $T^2$. The local temperature gradient also induces strain which can be responsible for shifting the QD emission. To release the strain in the suspanded membrane, we used a focused laser beam to cut some of the connection bridges next to the QD. After the strain release we still observed the same shift of the QD with the heating power which indicates that the shift is mainly due to temperature.

	Not only the QDs but also the PC cavities shift their resonant frequency with temperature. The local heating technique was used to shift a PC cavity located on the $w = 320 nm$ structure.  Using the same heating power as for the QD tuning, we observed the cavity resonance red shift by up to 0.48 nm (Fig. 3 (a) and (b)), about 3 times less than the QD shift. The quality factor of the cavity dropped from 7600 to 4900 because of the material loss increase at higher temperatures. Beside temperature tuning, photonic crystal cavities can be tuned using chemical digital etching \cite{HennessyDigitalEtch} or by depositon of molecular layers on top of the PC membrane \cite{StraufMonolayerCavityTuning} .

\begin{figure}[htbp]
    \includegraphics[width=3.375in]{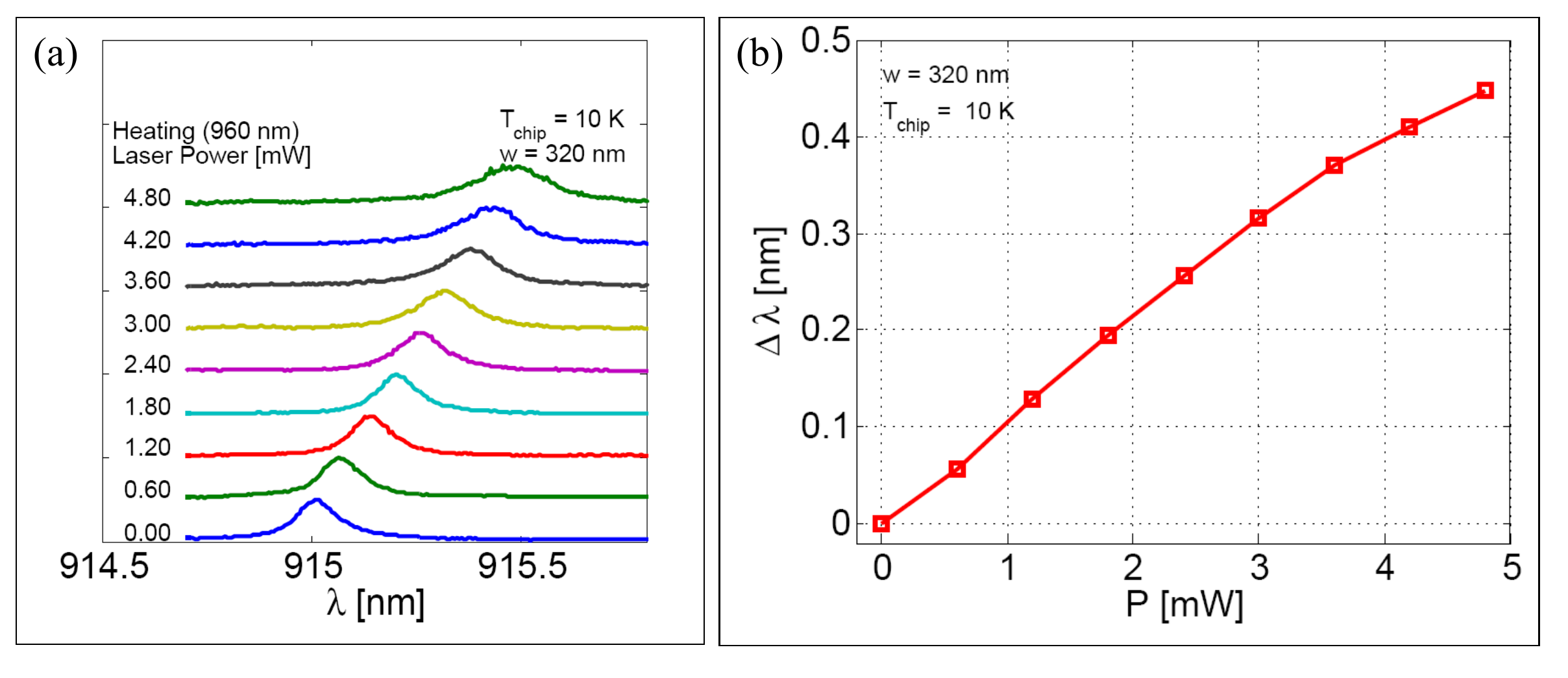}
    \caption{(a) Detuning of the PC cavity resonance with increasing temperature due to local heating. (b) Dependence of the PC cavity resonance wavelength on the local heating power. }
    \label{fig3}
\end{figure}

	A problem of immediate concern in cavity-QED experiments is spectral tuning of the QD onto the cavity resonance. We used our technique to locally tune two quantum dot lines into a photonic crystal cavity mode with $Q = 9000$, as shown in Fig. 4. As they enter the cavity spectrum, the continuously-pumped QDs experience Purcell-enhanced spontaneous emission rate, leading to higher emission (Fig.4 (b)). Though strong coupling is possible with this type of design, in this case we only observe moderate emission rate enhancement due to significant spatial mismatch between QDs and cavity mode.
	
	\begin{figure}[htbp]
    \includegraphics[width=3.375in]{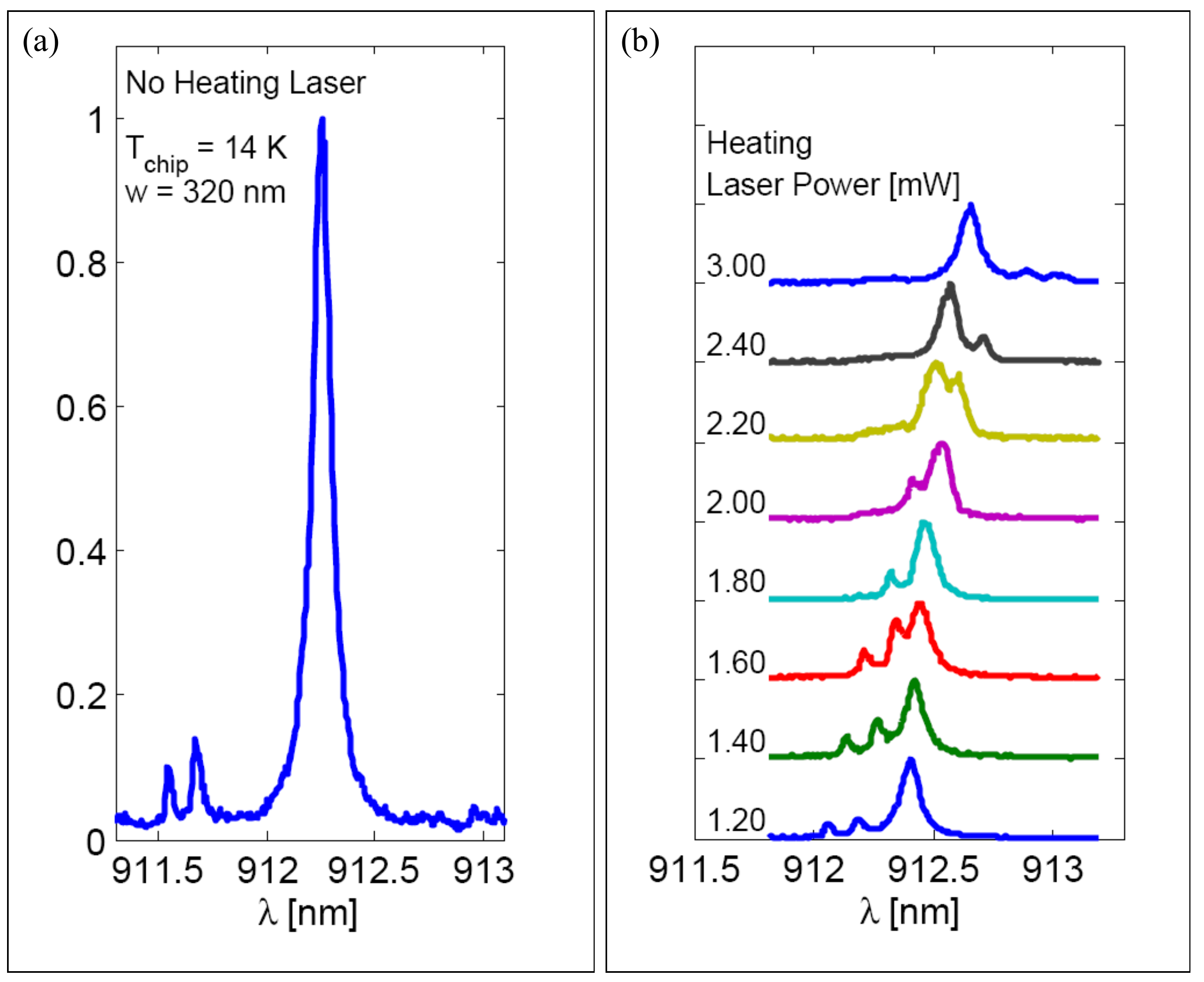}
    \caption{(a) Spectrum showing a PC cavity and two QDs detuned from the cavity resonance. (b) Spectra showing the tuning of the QDs into resonance with the cavity mode using the local heating technique.}
    \label{fig4}
\end{figure}

	The proof of concept experiment presented in our paper shows the local tuning of PC cavities and QDs that are not integrated into a PC circuit. However, it is only a matter of design to build PC circuits with integrated local heaters that could enable the independent tuning of different components of the same circuit. Moreover, this technique can be slightly modified by changing the heating method from optical to electrical by connecting electrical wires to the metal pads.

	Our local temperature tuning technique is completely reversible and does not affect the structure of the PC cavities or the QDs. Another tuning technique that relies on locally heating microcavities to permanently change the structure of the resonator and the QDs has been reported by Rastelli et. al. \cite{InSituLaserProcRastelli}.
	 
  In conclusion, we have demonstrated a novel technique for in-situ tuning of QDs by up to $1.4 nm$ without significant deterioration in the QD emission. This method works locally and reversibly, making it a useful tool for a range of solid state studies, from local thermometry to quantum information science. In particular, the method is compatible with photonic crystal structures and forms a crucial step towards building an on-chip quantum network involving resonant QDs.

	Financial support was provided by the MURI Center for photonic quantum information systems (ARO/DTO program No. DAAD19-03-1-0199), ONR Young Investigator Award and NSF Grant No. CCF-0507295. Work was performed in part at the Stanford Nanofabrication Facility of NNIN supported by the National Science Foundation under Grant ECS-9731293.



\newpage

\begin{thebibliography}{10}

\bibitem{05PRLEnglund}
D.~Englund, D.~Fattal, E.~Waks, G.~Solomon, B.~Zhang, T.~Nakaoka, Y.~Arakawa,
  Y.~Yamamoto, and J.~Vu\v{c}kovi\'{c}.
\newblock {Controlling the Spontaneous Emission Rate of Single Quantum Dots in
  a Two-Dimensional Photonic Crystal}.
\newblock {\em Physical Review Letters}, 95(013904), July 2005.

\bibitem{Yoshie04}
T.~Yoshie, A.~Scherer, J.~Hendrickson, G.~Khitrova, H.~M. Gibbs, G.~Rupper,
  C.~Ell, O.~B. Shchekin, and D.~G. Deppe.
\newblock {Vacuum Rabi splitting with a single quantum dot in a photonic
  crystal nanocavity}.
\newblock {\em Nature}, 432:200--203, November 2004.

\bibitem{CZKM1997PRL}
J.~I. Cirac, P.~Zoller, H.~J. Kimble, and H.~Mabuchi.
\newblock {Quantum State Transfer and Entanglement Distribution among Distant
  Nodes in a Quantum Network}.
\newblock {\em Physical Review Letters}, 78(16):3221--24, April 1997.

\bibitem{DirkAndreiSPhtransfer}
D.~Englund, A.~Faraon, B.~Zhang, Y.~Yamamoto, and J.~Vu\v{c}kovi\'{c}.
\newblock {Generation and Transfer of Single Photons on a Photonic Crystal
  Chip}.
\newblock {\em ArXiv}, quant-ph(0609053), Sep 2006.

\bibitem{HennessyDeterministicCoupling}
A.~Badolato, K.~Hennessy, M.~Atat{\"u}re, J.~Dreiser, E.~Hu, P.M. Petroff, and
  A.~Imamoglu.
\newblock {Deterministic Coupling of Single Quantum Dots to Single Nanocavity
  Modes}.
\newblock {\em Science}, 308(5725):1158 -- 1161, May 2005.

\bibitem{PhotolithQDRegistration}
K.H. Lee, A.M. Green, R.A. Taylor, D.N. Sharp, J.~Scrimgeour, O.M. Roche, J.H.
  Na, A.F. Jarjour, A.J. Turberfield, F.S.F. Brossard, D.A. Williams, and
  G.A.D. Briggs.
\newblock {Registration of single quantum dots using cryogenic laser
  photolithography}.
\newblock {\em Applied Physics Letters}, 88(193106), 2006.

\bibitem{SCImamogluNature}
K.~Hennessy, A.~Badolato, M.~Winger, D.~Gerace, M.~Atat{\"u}re, S.~Gulde,
  S.~Falt, E.L. Hu, and A.~Imamoglu.
\newblock {Quantum nature of a strongly coupled single quantum dot-cavity
  system}.
\newblock {\em Nature}, 445:896--899, Feb 2007.

\bibitem{StarkKarraiPRL}
A.~H{\"o}gele, S.~Seidl, M.~Kroner, K.~Karrai, R.J. Warburton, B.D. Gerardot,
  and P.M. Petroff.
\newblock {Voltage-Controlled Optics of a Quantum Dot}.
\newblock {\em Physical Review Letters}, 93(217401), Nov 2004.

\bibitem{ZeemanKarraiPRL}
D.~Haft, C.~Schulhauser, A.Q. Govorov, R.J. Warburton, K.~Karrai, J.M. Garcia,
  W.~Schoedfeld, and P.M. Petroff.
\newblock {Magneto-optical properties of ring-shaped self-assembled InGaAs
  quantum dots}.
\newblock {\em Physica E}, 13:165--169, 2002.

\bibitem{TTunePetroffAPL}
A.~Kiraz, P.~Michler, C.~Becher, B.~Gayral, A.~Imamoglu, L.~Zhang, E.~Hu, W.V.
  Schoenfeld, and P.M. Petroff.
\newblock {Cavity-quantum electrodynamics using a single InAs quantum dot in a
  microdisk structure}.
\newblock {\em Applied Physics letters}, 78(25):3932--3934, June 2001.

\bibitem{InGaAsStrainTune}
Stefan Seidl, Martin~Kroner amd Alexander~H{\"o}gele, Khaled Karrai, Richard~J.
  Warburton, Antonio Badolato, and Pierre~M. Petroff.
\newblock {Effect of uniaxial stress on excitons in a self-assembled quantum
  dot}.
\newblock {\em Applied Physics letters}, 88(203113), 2006.

\bibitem{FonGaAsThermalCond}
W.~Fon, K.C. Schwab, J.M. Worlock, and M.L. Roukes.
\newblock {Phonon scattering mechanisms in suspended nanostructures from 4 to
  40 K}.
\newblock {\em Physical Review B}, 66(045302), 2002.

\bibitem{GaAsBandGapThurmond}
C.D. Thurmond.
\newblock {The standard thermodynamic functions for the formation of electrons
  and holes in Ge, Si, GaAs and GaP}.
\newblock {\em J. Electrochem. Soc.}, 122(1133), 1975.

\bibitem{HennessyDigitalEtch}
K.~Hennessy, A.~Badolato, A.~Tamboli, P.~M. Petroff, E.~Hu, M.~Atat{\"u}re,
  J.~Dreiser, and A.~Imamoglu.
\newblock {Tuning photonic crystal nanocavity modes by wet chemical digital
  etching}.
\newblock {\em Applied Physics Letters}, 87(021108), 2005.

\bibitem{StraufMonolayerCavityTuning}
S.~Strauf, M.~T. Rakher, I.~Carmeli, K.~Hennessy, C.~Meier, A.~Badolato,
  M.~J.~A. DeDood, P.~M. Petroff, E.~L. Hu, E.~G. Gwinn, and D.~Bouwmeester.
\newblock {Frequency control of photonic crystal membrane resonators by
  monolayer deposition}.
\newblock {\em Applied Physics Letters}, 88(043116), 2006.

\bibitem{InSituLaserProcRastelli}
A.~Rastelli, A.~Ulhaq, S.~Kiravittaya, L.~Wang, A.~Zrenner, and O.G. Schmidt.
\newblock {In situ laser microprocessing of single self-assembled quantum dots
  and optical microcavities}.
\newblock {\em Applied Physics Letters}, 90(073120), 2007.

\end{thebibliography}

\end{document}